\documentclass[
    ,final            
  ]
  {aipproc}

\layoutstyle{6x9}

\newcommand{\ltsima}{$\; \buildrel < \over \sim \;$}
\newcommand{\lsim}{\lower.5ex\hbox{\ltsima}}
\newcommand{\gtsima}{$\; \buildrel > \over \sim \;$}
\newcommand{\gsim}{\lower.5ex\hbox{\gtsima}}
\newcommand{\be}{\begin{equation}}
\newcommand{\ee}{\end{equation}}

\newcommand{\AmS}{{\protect\the\textfont2
  A\kern-.1667em\lower.5ex\hbox{M}\kern-.125emS}}

\def\etamin{\eta_*}
\def\dec{np}
\def\no{\noindent}
\begin{document}

\title{Heating and Deceleration of GRB Fireballs by Neutron Decay  }

\author{Elena  M. Rossi}{
  address={Institute of Astronomy, University of Cambridge, Madingley Road,
Cambridge CB3 0HA, England }
}

\author{Andrei M. Beloborodov }{
  address={Physics Department,
        Columbia University, 538 W 120th Street, New York, NY 10027, USA }
}
\author{Martin J. Rees}{
  address={Institute of Astronomy, University of Cambridge, Madingley Road,
Cambridge CB3 0HA, England }
}

\begin{abstract}
Fireballs with high energy per baryon rest mass ($\eta>\eta_*\sim
400$) contain a relatively slow neutron component. We show here that
in this situation the thermal history of fireballs is very different
from the standard adiabatic cooling.

\end{abstract}

\maketitle


%
%

\section{Introduction}

Baryonic matter ejected in a GRB explosion is partially composed of
free neutrons (see \citep{Der} and \citep{B03a} for detailed
calculations and references).  This has a strong impact on the
external blast wave and changes the afterglow mechanism at radii
$r\lsim 10^{17}$~cm \citep{B03b}.  We here investigate the dynamical
effect of neutrons on the early evolution of a relativistic fireball,
prior to the afterglow phase.

\section{Fireball acceleration and neutron-proton decoupling}

At the initial stage of the fireball acceleration, the neutrons are
collisionally coupled to the protons and accelerated by radiation
pressure at the same rate. If the energy-per-baryon, $\eta$, is high
enough, neutrons decouple before the acceleration stage is completed.
This happens if

\begin{equation}
\eta\ge\etamin\simeq \left\{ \begin{array}{ll}

 400\,\left[\frac{L_{52}} {R_{07}\frac{1+\xi}{2}} \right]^{1/4} & \xi \le 1\\
 400\,\left[\frac{L_{52}\,\xi} {R_{07}\frac{1+\xi}{2}} \right]^{1/4} & \xi > 1
	\end{array}\right.
\label{eq:eta*}
\end{equation}

\noindent
where $\xi$ is the ratio of neutron and proton densities,
$R_{0}=10^{7}\,{\rm cm}\, R_{07}$ is the initial size of the fireball, and
$L=10^{52}\,{\rm erg}\,L_{52}$ is the equivalent isotropic luminosity
of the explosion. Beyond the decoupling radius,
\be
R_{\dec}\simeq R_0\,\eta_*\,\left(\frac{\eta_*}{\eta}\right)^{1/3},
\label{eq:rdec}
\ee
neutrons coast with a constant Lorentz factor

\be
\Gamma_{n}\simeq \frac{R_{np}}{R_0}\simeq\,\, \eta_* \,\left(\frac{\eta_*}{\eta}\right)^{1/3}.
\label{eq:gn}
\ee


The optically thick proton fireball continues to accelerate as
$\Gamma_p\simeq r/R_0$ until the internal energy equals the proton
rest-mass energy and acceleration is no longer efficient. If the
fireball remains optically thick during the whole acceleration stage,
its Lorentz factor saturates at

\noindent
 \be
\Gamma_p \simeq \left\{ \begin{array}{ll}
 \eta & \eta < \etamin, \\
 \eta \left(1+\xi-\xi\,\left(\frac{\eta_{*}}{\eta}\right)^{4/3}\right)
                                                   & \eta>\etamin,
	\end{array}\right.
\label{eq:gs}
\ee
and the acceleration stage ends at radius
\be
R_s \simeq \Gamma_p R_0.
\label{eq:rs}
\ee


\section{Adiabatic cooling}

\noindent
We first consider the thermal history of the fireball without neutrons.
Electrons, protons, and radiation maintain a common
temperature $T$ in the early dense fireball via Coulomb collisions and
Compton scattering, and $T$ decreases adiabatically to very low
values.  During such adiabatic expansion, the photon-to-baryon ratio
$\phi=n_\gamma/n_b=const\sim 10^5$, and radiation completely dominates
the internal energy and pressure.  Therefore, $T$ decreases according
to the radiation adiabatic law with index
$\hat{\gamma}=\frac{4}{3}$. This continues until the fireball becomes
transparent to radiation at the photosphere radius $R_{\tau}$.
Assuming $R_\tau>R_s$, we have

\be
R_{\tau}= \frac{L \sigma_T\,} {4\pi m_p c^3 \Gamma_p^3},
\label{eq:rt}
\ee

\noindent
here $\sigma_T$ is Thomson cross section.  After becoming transparent
the plasma is still tracking the temperature of the (freely streaming)
photons, which is constant in the plasma frame if the fireball coasts
with a constant Lorentz factor. The electrons decouple thermally from
radiation only when the Compton timescale
($t_C=\Gamma_p(3m_ec/4\,U_{rad}\sigma_T)$ where $U_{rad}$ is the
radiation energy density) exceeds the expansion timescale $R/c$.  This
happens at radius

\be
R_{e\gamma}=\frac{\sigma_T L}{3\pi\,m_ec^3\,\Gamma_p^{7/3}}\,
  \left(\frac{R_0}{R_{\tau}}\right)^{2/3}.
\label{eq:ref}
\ee

\no
The protons decouple from the electrons when the Coulomb timescale
$t_{ep}\approx 17\,T^{3/2}n_e^{-1}\Gamma_p$~s exceeds $R/c$. 
The corresponding radius of e-p decoupling is

\be
R_{ep}\simeq\left(68\pi\,m_pc^4\right)^{-1}
\frac{L}{\Gamma_p^{3}\,T_s^{3/2}}\left(\frac{ R_{\tau}} {R_s}\right),
\ee
 
\noindent
where $T_s=T_0/\Gamma_p $  is the temperature at the saturation
radius, and $T_0$ is the initial temperature of the fireball. The
protons decouple thermally from the electrons before the electrons
decouple from radiation if $R_{ep}<R_{e\gamma}$. One can show that
$R_{ep}/R_{e\gamma}\approx 4 \times
10^{-2}T_{s,7}^{-3/2}(R_\tau/R_s)^{5/3}$ (with $T_s=T_{s7} \times
10^{7}$~K).  At radii $R>\min{(R_{e\gamma},R_{ep})}$, the protons are
decoupled from radiation and cool adiabatically with index
$\hat{\gamma}=\frac{5}{3}$.

\begin{figure}
\centerline{\resizebox{0.5\textwidth}{!}{\includegraphics*{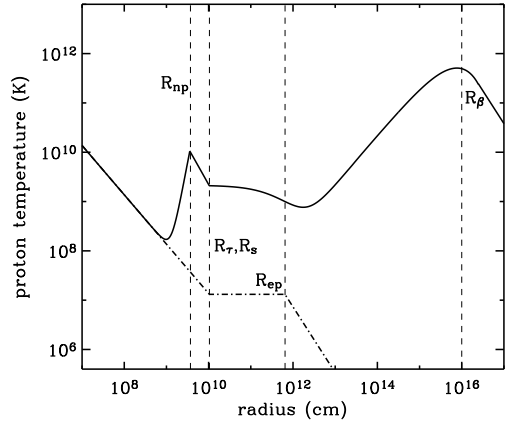}}}
\caption{Proton temperature as a function of radius in a fireball with 
neutron-to-proton ratio $\xi=1$ (solid curve).  The relevant radii
$R_{ep}$, $R_{np}$, $R_\tau$, $R_s$, and $R_\beta$ are
shown by vertical dashed lines (note that $R_s=R_{\tau}$ in this
case). For comparison, the dot--dashed curve shows the evolution of a
pure proton fireball ($\xi=0$) with the same luminosity $L=10^{52}$
erg s$^{-1}$, initial radius $R_0=10^{7}$ cm and final Lorentz factor
$\Gamma_p\simeq 1040$.}
\label{fig:tnp}
\end{figure}

Thus, in the absence of neutrons, the fireball cools down
adiabatically as $T\propto r^{-1}$ during the acceleration stage and
as $T\propto r^{-2/3}$ during the subsequent coasting stage ($\Gamma_p
\sim \eta$) up to the transparency radius.
 After thermal decoupling from radiation the protons cool as
 $T_p\propto r^{-4/3}$. The thermal history of a pure proton fireball
 is shown by dot-dashed curve in Fig.~\ref{fig:tnp}.  For most of the
 early evolution, the fireball is a cold coasting outflow.  The
 presence of neutrons and their $\beta$-decay change this picture.

\section{Heating by proton--neutron collisions}

Protons in the fireball are continuously heated via proton-neutron
collisions \citep{p01}. The collisional heating reaches its peak at
$R\approx R_{np}$: at $R<R_{np}$ the n-p collisions are frequent but
the relative velocity of the neutron and proton components is small
($\beta_{np}\approx\frac{t_{np}}{R/c}<1$) and at $R>R_{np}$ the
relative velocity is relativistic $\beta_{np}\sim 1$ however the
collisions are rare.  The collisional heating, thus, peaks at $R_{np}$
where the relative velocity becomes comparable to the speed of light
and the rate of collisions still allows an efficient transfer of
energy on a dynamical timescale.  There are two sinks of heat gained
by protons via n-p collisions: adiabatic cooling and Coulomb
scattering off electrons.  The competition between heating and cooling
shapes the first peak in the proton temperature profile (Fig.~1).  At
$R=R_{ep}$ the energy transfer from the protons to the electrons
becomes inefficient on the expansion timescale $R/c$.  Then most of
the heat gain by protons remains stored in the proton component, not
given to the electrons and radiation.

\section{Fireball heating and deceleration by decayed neutrons}

In a fireball with $\eta>\etamin$, the neutrons have a smaller
momentum than the protons after $R_{np}$. Their decay products $e^-$
and $p$ exchange momentum with the ion fireball and heat it up.  The
heating of an ion medium by $\beta$-decay of neutrons moving with
respect to the medium was discussed in [3].  The decay particles
exchange momentum with the medium through the two-stream instability
or because of gyration in a transverse magnetic field frozen into the
medium. Since the neutrons decay gradually at all radii, this momentum
exchange and heating take place continuously from the moment of
decoupling.

The heating rate of the fireball due to $\beta$-decay is
\be
\frac{dE_{h}}{dr}= (\Gamma_{rel}-1)\,\frac{dM_{np}c^2}{dr},
\label{eq:dEh}
\ee
where $\Gamma_{rel}$ is the neutron Lorentz factor in the fireball
rest frame,
\be
\frac{dM_{np}}{dr}=\frac{M_n}{R_{\beta}},
\ee
is the decayed neutron mass per unit radius, and
\be
R_{\beta}\simeq \,0.8\times 10^{16}\,\, {\rm cm}\,\, \left(\frac{\Gamma_{n}}{300}\right)
\ee
is the mean radius of decay.
The heating by $\beta$-decay begins at $R_{\rm{\dec}}$ and
significantly changes the thermal history of the fireball. It shapes
the second peak in proton temperature (see Fig.~\ref{fig:tnp}).

As long as proton thermal energy is much below
$m_pc^2$, adiabatic cooling is small compared to heating and 
the proton temperature increases linearly with radius,
$$
T_p \propto r.
$$ 
At a radius $R_{dis}$ the dissipated bulk kinetic energy
equals the total energy of the fireball and $\Gamma_p$ begins to
decrease.  Between $R_{dis}$ and $R_{\beta}$ the
$\beta$-decay heating balances the adiabatic cooling and 
$$
\Gamma_p \propto\left(\frac{R_{\beta}}{r}\right)^{1/2}.
$$
When the fireball
reaches $R_{\beta}$ the neutron component is exponentially depleted
and the dissipation process switches off. The adiabatic cooling then
leads to a power-law decrease in temperature,  
$$
T_{p} \propto r^{-4/3}.
$$

\noindent

\section{Discussion}
We have shown here that the presence of a
neutron component greatly affects the early dynamics of GRB fireballs
with $\eta>\eta_*\sim 400$. Contrary to the pure proton model, 
the fireball with a neutron component heats up significantly at 
$R\gsim R_{np}$.
In a popular GRB scenario, the bulk energy is partially converted into 
the observed $\gamma-$rays through shock waves inside
the fireball at $r\gsim 10^{12}$~cm \citep{RM94}. 
The $\beta$-decay process described here 
should affect the internal shocks. It decelerates
the fastest portions
of the inhomogeneous fireball and reduces the contrast of Lorentz
factors, which should significantly reduce the dissipation efficiency 
of internal shocks. A detailed study of these effects is in preparation.

%
%

\begin{figure}
\centerline{\resizebox{0.5\textwidth}{!}{\includegraphics*{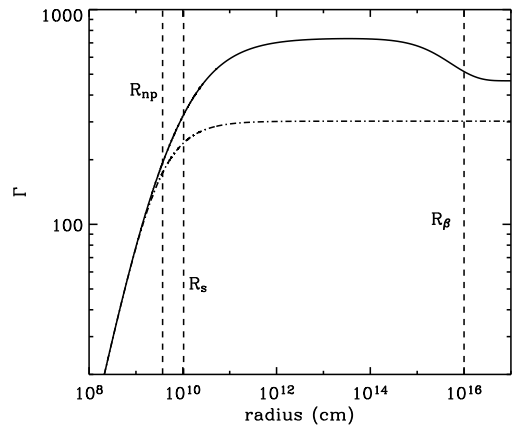}}}
\caption{The bulk Lorentz factor of protons (solid line) and neutrons 
(dot-dashed line) as a function of radius for the same parameters as 
in Fig.~\ref{fig:tnp}.
}
\label{fig:gnp}
\end{figure}



\bibliographystyle{aipproc}   

\bibliography{rossi_elena_0-rev.2}

\IfFileExists{\jobname.bbl}{}
 {\typeout{}
  \typeout{******************************************}
  \typeout{** Please run "bibtex \jobname" to obtain}
  \typeout{** the bibliography and then re-run LaTeX}
  \typeout{** twice to fix the references!}
  \typeout{******************************************}
  \typeout{}
 }

\end{document}